\pdfoutput=1
\documentclass[twocolumn,preprint]{aastex62}
\usepackage[utf8x]{inputenc}
\usepackage{amsmath,amsfonts,amssymb,mathrsfs,bm,enumerate}
\usepackage[T1]{fontenc}
\usepackage[all]{hypcap}
\hypersetup{breaklinks,colorlinks,citecolor=blue,linkcolor=magenta}

\shorttitle{Smoothly creating GJ876}
\shortauthors{Dempsey \& Nelson}

\begin{document}

\title{Forming Gliese 876 Through Smooth Disk Migration}
\author{Adam M. Dempsey}
\affiliation{Center for Interdisciplinary Exploration and Research in Astrophysics (CIERA)
and
Department of Physics and Astronomy
Northwestern University \\
2145 Sheridan Road
Evanston, IL 60208
USA}
\author{Benjamin E. Nelson}
\affiliation{Center for Interdisciplinary Exploration and Research in Astrophysics (CIERA)
and
Department of Physics and Astronomy
Northwestern University \\
2145 Sheridan Road
Evanston, IL 60208
USA}
\affiliation{Northwestern Institute for Complex Systems \\
600 Foster Street
Evanston, IL 60208
USA}

\email{adamdempsey2012@u.northwestern.edu}

\begin{abstract}
We run a suite of dissipative N-body simulations to determine which regions of phase space for smooth disk migration are consistent with the GJ876 system, an M-dwarf hosting three planets orbiting in a chaotic 4:2:1 Laplace resonance. 
We adopt adaptive mesh refinement (AMR) methods which are commonly used in hydrodynamical simulations to efficiently explore the parameter space defined by the semi-major axis and eccentricity damping timescales. 
We find that there is a large region of phase space which produces systems in the chaotic Laplace resonance and a smaller region consistent with the observed eccentricities and libration amplitudes for the resonant angles. 
Under the assumptions of Type I migration for the outer planet, we translate these damping timescales into constraints on the protoplanetary disk surface density and thickness. 
When we strongly (weakly) damp the eccentricities of the inner two Laplace planets, these timescales correspond to disk surface densities around ten thousand (a few hundred) grams per square centimeter and disk aspect ratios between $1-10\%$. 
Additionally, smooth migration produces systems with a range of chaotic timescales, from decades and centuries to upwards of thousands of years. 
In agreement with previous studies, the less chaotic regions of phase space coincide with the system being in a low energy double apsidal corotation resonance.
Our detailed modeling of multi-planetary systems coupled with our AMR exploration method enhances our ability to map out the parameter space of planet formation models, and is well suited to study other resonant chain systems such as Trappist-1, Kepler-60, and others.

\end{abstract}

\keywords{celestial mechanics  --- planet-disk interactions  --- planets and satellites: dynamical evolution and stability}

\section{Introduction} \label{sec:1}

Gliese 876 (GJ876) is an M-dwarf star ($M_\star \simeq 0.37 M_\odot$) 4.6 pc away that harbors a well-studied multi-planetary system \citep{1998ApJ...505L.147M,2007ASSL..350.....V,2014MNRAS.438.2413V}.
Through radial velocity measurements, we know of at least four planets orbiting GJ876, the outer three of which are in a 4:2:1 (Laplace) mean-motion resonance (MMR) \citep{2010ApJ...719..890R}. 
Leveraging the strong resonant interactions, several studies have been able to break the mass and inclination degeneracy of the system, providing strong constraints on the full orbital parameters and planetary masses \citep{2010A&A...511A..21C,2010ApJ...719..890R,2016MNRAS.455.2484N,2018A&A...609A.117T,2018AJ....155..106M}.
Of particular note is that the system is chaotic on potentially observable timescales, and that the mass ratios of GJ876-c ($M_c \simeq 265 M_\oplus$, $P_c = 30 d$, $e_c = 0.26$) and GJ876-b ($M_b \simeq 845 M_\oplus$, $P_b = 61 d$, $e_b = 0.036$) are large enough to potentially open a deep and wide gap in their natal disk -- possibly extending out to the outermost planet, GJ876-e ($M_e \simeq 16 M_\oplus$, $P_e = 124 d$, $e_e \simeq  0.03$) \citep{2001A&A...374.1092S,2010A&A...510A...4R}.

Recently, \cite{2015AJ....149..167B} investigated the formation of GJ876 through \emph{stochastic} disk migration.
They attributed the stochasticity to turbulence in the disk, and argued that it was essential for producing the observed chaotic state of the system.
Historically, the magnetorotational instability (MRI) was thought to be the driver of turbulence in protoplanetary disks \citep{1991ApJ...376..214B}. 
However, recent simulations have shown that non-ideal MHD processes can suppress or completely shutoff the MRI in the midplanes of protoplanetary disks where planets are expected to form \citep[e.g.][]{2013ApJ...769...76B,2014A&A...566A..56L,2015MNRAS.454.1117S}. 
If these regions are laminar or weakly turbulent, then stochastic forcing of a smoothly migrating planet may be negligible.
It then becomes important to know whether or not smooth migration alone can account for all of the observed characteristics of GJ876.
Moreover, while previous studies have focused on constructing the 2:1 resonance of GJ876-c,b through smooth migration \citep[e.g.,][]{2002ApJ...567..596L}, there has yet to be an exhaustive analysis focused on constructing the chaotic Laplace resonance through disk migration.

This paper is organized as follows.
In \S\ref{sec:2} we outline our procedure for simulating the migration history of GJ876 and our method for efficiently exploring parameter space. 
In \S\ref{sec:3} we present the main results from our suite of simulations.
In \S\ref{sec:4} we explore several different variations on the procedure outlined in \S\ref{sec:2}. 
Finally, we end in \S\ref{sec:5} by discussing improvements to our smooth migration model and the possibility of extending our procedure to other resonant chain exoplanet systems.

\section{Numerical model} \label{sec:2}
\subsection{Smooth migration}

Smooth migration models attempt to parameterize the complex energy and angular momentum transfer processes between a planet and its surrounding disk by specifying the exponential damping timescales associated with semi-major axis evolution, $\tau_a \equiv | a/\dot{a}|$, and eccentricity evolution, $\tau_e \equiv |e/\dot{e}|$ \citep[for a review see e.g.][]{2012ARA&A..50..211K}.
The damping timescales for a planet of mass $M_p$ orbiting a star of mass $M_\star$ at semi-major axis $a$ and orbital period $P$ are, 
\begin{align} 
    \tau_a^{-1} = C_a  \left(\frac{4 \pi a^2 \Sigma_p M_p}{M_\star^2}\right)  \left(\frac{H}{a} \right)^{-2} P^{-1} , \label{eq:tau_a} \\
    \tau_e^{-1} = C_e \left(\frac{4 \pi a^2 \Sigma_p M_p}{M_\star^2}\right) \left(\frac{H}{a} \right)^{-4} P^{-1} , \label{eq:tau_e}
\end{align}
where $\Sigma_p$ and $H/a$ are the surface density and aspect ratio of the disk at the location of the migrating planet \citep{1980ApJ...241..425G}.
We adopt the proportionality constants $C_a = 2.175$ and $C_e = 0.39$ \citep{2002ApJ...565.1257T,2004ApJ...602..388T}, but note that they  depend strongly on the local disk structure \citep[e.g.][]{1980ApJ...241..425G,1981ApJ...243.1062G,1993ApJ...419..155A,1993ApJ...419..166A,2003ApJ...585.1024G,2012ARA&A..50..211K}.
Equations \eqref{eq:tau_a} and \eqref{eq:tau_e} correspond to the so-called Type I migration regime \citep{1997Icar..126..261W,2014AJ....147...32G}.
The ratio of these timescales, $K$, depends on the local disk aspect ratio as,
\begin{equation} \label{eq:K}
    K \equiv \frac{\tau_a}{\tau_e} \propto \left( \frac{H}{a} \right)^{-2}.
\end{equation}

In the N-body simulations presented in the next section, we only apply semi-major axis damping forces to GJ876-e as its migration rate should be relatively fast compared to the slow Type II migration rates of GJ876-c,b \citep[e.g.][]{2012ARA&A..50..211K,2015A&A...574A..52D}. 
However, since the eccentricity damping rates of the inner planets may be faster than their semi-major axis damping rates by a factor of $K$, and given that the equilibrium eccentricities of planets migrating in MMR depend on their eccentricity damping timescale \citep[e.g.][]{2002ApJ...567..596L,2014AJ....147...32G}, we apply eccentricity damping forces to the planets GJ876-c,b with a damping timescale denoted by $\tau_{e,1}$, in our N-body simulations.

\subsection{Simulation setup} \label{sec:setup}
Our goal is to map out the parameter space in $K$, $\tau_a$, and $\tau_{e,1}$ for the system properties of GJ876.  
Our simulation procedure is similar to that of \citet{2017ApJ...840L..19T}, who studied the formation history of the Trappist-1 resonant chain. 
For a given $K$, $\tau_a$, and $\tau_{e,1}$ we first initialize a coplanar, three planet system around a $M_\star = 1 M_\odot$ star. 
The masses of the three planets are chosen to reproduce the best fit mass ratios of planets GJ876-c,b and e (which we henceforth label as planets 1, 2, and 3).
These are set to $M_1 = 2.15\times10^{-3} M_\star$, $M_2 = 6.85 \times 10^{-3} M_\star$, and $M_3 = 1.30 \times 10^{-4} M_\star$ \citep{2018AJ....155..106M}.
The initial orbital periods are $P_1 = 1$ year, $P_2/P_1= 2.2$, and $P_3/P_2 = 10$.
Once initialized, we integrate each planetary system for a time of $10 \tau_a$ while applying the aforementioned damping forces.

After this initial damping phase, we remove all damping forces from the system over a time of $\tau_a$. 
If the system is in the Laplace resonance, we rescale the system to the observed properties of GJ876.
This entails changing the mass of the central star to $M_\star = 0.37 M_\odot$ and the inner planet's period to the observed period of GJ876-c, all while keeping the planet-to-star mass ratios and period ratios the same as in the previous phase.
Finally, we integrate the rescaled system without any damping forces for $10^5$ years.
This rescaling of the simulation after the migration phase is similar to what was done in \citet{2017ApJ...840L..19T} for the Trappist-1 system, and allows us to avoid having to fine tune the initial conditions or adjust the duration of the damping phase so that the planets end up near their observed periods by the end of the damping period.

All integrations are done with the REBOUND\footnote{\url{http://github.com/hannorein/rebound}, v3.5.2.} N-body code \citep{Rein:2012cd}. 
For the damping forces we use the REBOUND extension REBOUNDX\footnote{\url{http://github.com/dtamayo/reboundx}, v2.18.1.} to introduce orbit-averaged forces which give the proper damping timescales \citep{2000MNRAS.315..823P}.
We use the WHFast integrator for all simulations with $40$ timesteps per initial orbital period of the inner planet \citep{Rein:2015cl}. 
We found this to be in good agreement with simulations run with the more accurate (but slower) IAS15 integrator \citep{Rein:2015ib}.

\begin{figure}
    \centering
    \plotone{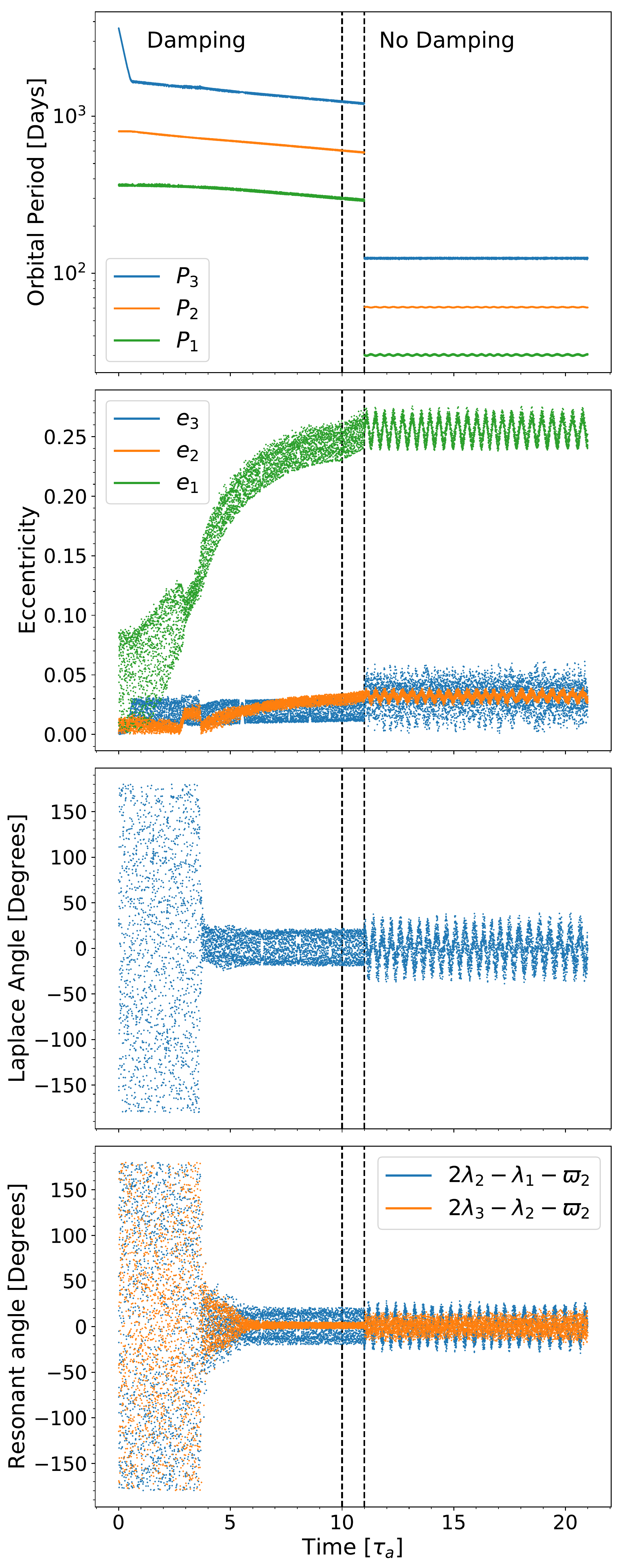} 
    \caption{Time evolution of a simulation with $(\log_{10}(K),\log_{10}(\tau_a/P_1),\log_{10}(\tau_{e,1}/P_1)) = (3.125,4,5.5)$ demonstrating our simulation procedure. The vertical lines mark the end of the initial damping phase and the start of the damping-free phase. During this time period the damping forces on the planets are reduced to zero over a time of $\tau_a$. At the beginning of the damping-free phase we rescale the system to the observed period of the inner planet. In descending order, the panels show the time evolution of the orbital periods, eccentricities, the Laplace resonant angle, and two, 2:1 MMR angles. }
    \label{fig:sim_demo}
\end{figure}

We show the time evolution of the planetary periods, eccentricities, Laplace resonance angle, and two 2:1 resonance angles of a representative simulation in Figure \ref{fig:sim_demo}.
The Laplace angle is defined in terms of the mean longitudes, $\lambda$, of the planets as $\Phi_L = \lambda_1 - 3 \lambda_2 + 2 \lambda_3$.
The 2:1 MMR resonant angles are defined as $\Phi_{1,2} = 2\lambda_2 - \lambda_1 - \varpi_2$ and $\Phi_{2,3} = 2 \lambda_3 - \lambda_2 - \varpi_2$, where $\varpi$ denotes the the longitude of pericenter. 
In this example simulation, for times less than $10 \tau_a$ the outer planet experiences damping forces with $ \log_{10} K = 3.125$ and $\tau_a = 10^4 P_1$. The inner planets also experience eccentricity damping forces with $\tau_{e,1} = 10^{5.5} P_1$. 
During this phase the planets are brought into successive 2:1 MMRs, evidenced by $\Phi_{1,2}$ and $\Phi_{2,3}$ librating about $0^\circ$ with small amplitude. 
At nearly the same time the system also catches into the Laplace resonance. 
As the system continues to migrate while in the 4:2:1 resonance, the eccentricity of the inner planet grows to $e_1 \approx 0.25$, while $e_2 \approx e_3 \approx 0.03$.
Between times of $10 \tau_a$ and $11 \tau_a$, which corresponds to the region between the two vertical dashed lines in Figure \ref{fig:sim_demo}, we gradually remove the damping forces.
Then at a time of $11 \tau_a$ we completely remove all damping forces and rescale the system to an inner period of $30$ days and a central mass of $0.37 M_\odot$. 
Finally, we evolve the system for $10^5$ years where it remains in a stable, resonant configuration. 

\begin{figure*}
    \centering
    \plotone{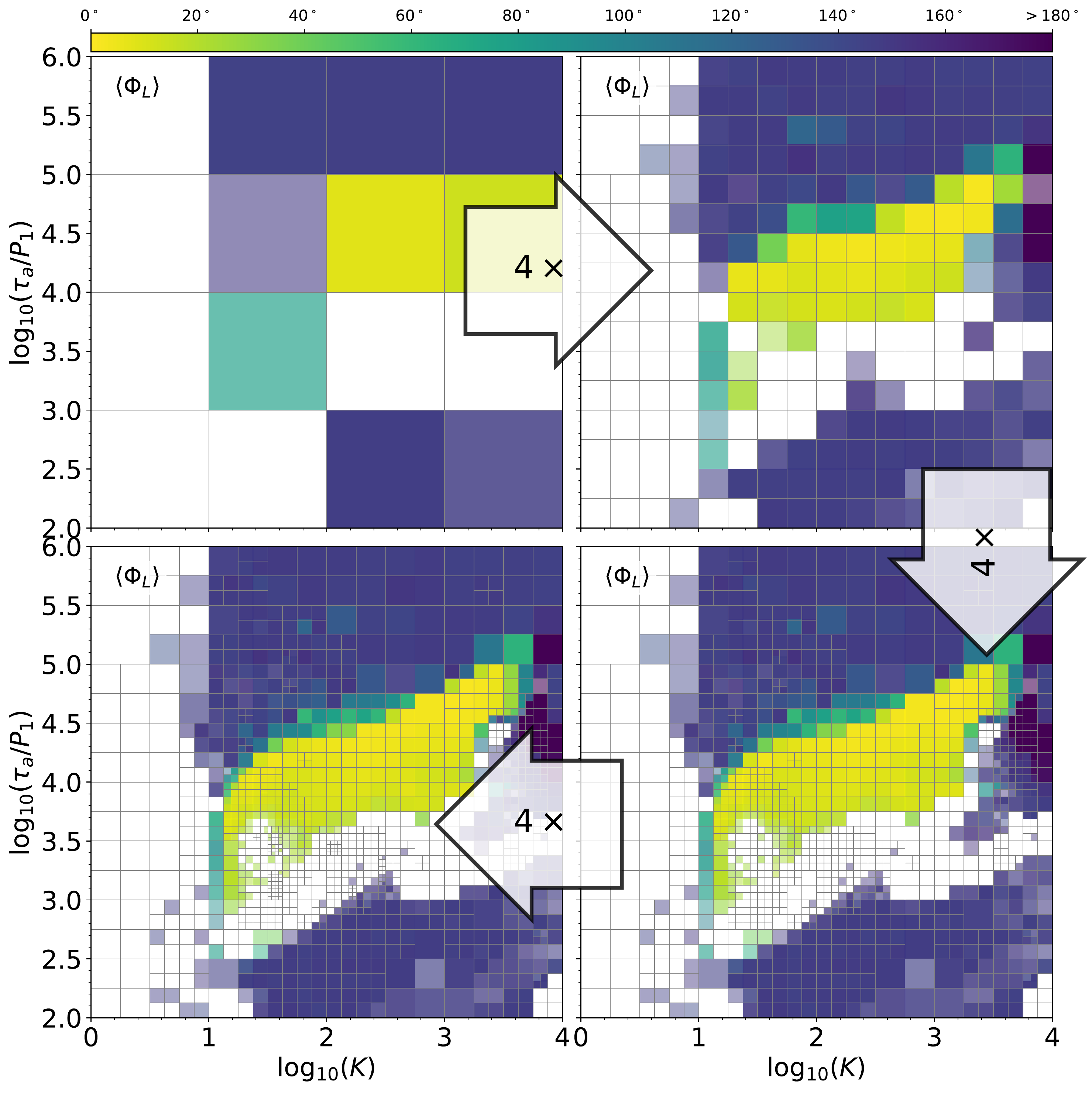}
    \caption{The Laplace angle libration amplitude and our AMR grid in $\tau_a$ and $K=\tau_a/\tau_e$ space for fixed $\tau_{e,1}=10^5 P_1$, at increasing levels of refinement. The color of each cell is the average libration amplitude of $10$ simulations where the initial angles were randomized. The opacity of each cell indicates its instability fraction, with white cells indicating that all $10$ simulations went unstable. The observed libration amplitude of GJ876 is approximately $26^\circ \pm 5^\circ$ \citep{2018AJ....155..106M}. }
    \label{fig:grid}
\end{figure*}

\subsection{Efficiently exploring parameter space}

\begin{figure*}[th]
    \centering
    \plotone{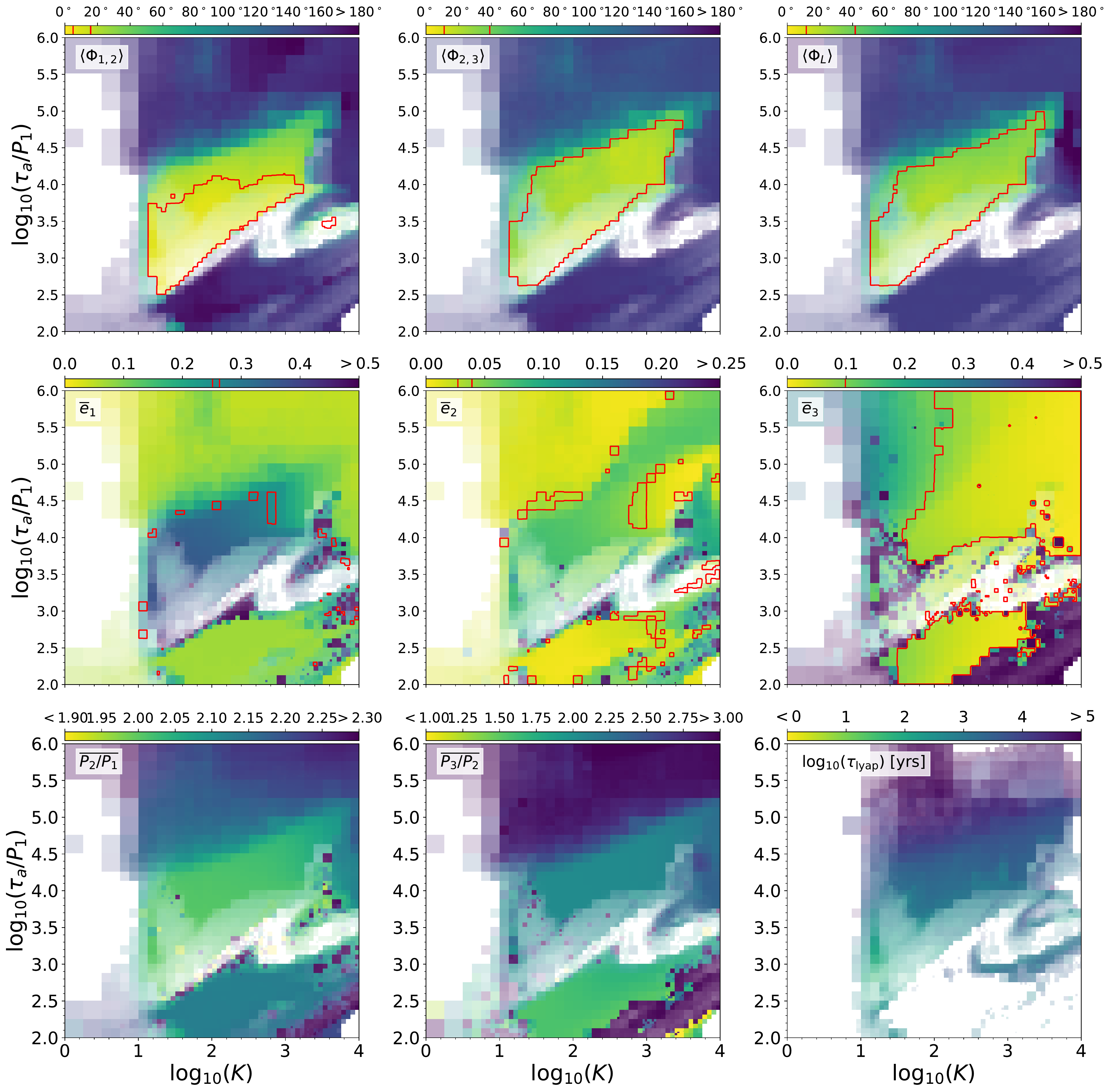}
    \caption{Overview of the resonant angles (top row), eccentricities (middle row), inner and outer period ratios, and Lyapunov time (bottom row) for the simulation procedure described in \S\ref{sec:setup}. All quantities are averaged over $\tau_{e,1}$.  An overbar denotes the time-averaged value, while angle brackets denote the RMS libration amplitude. When present, regions falling within the red contours signify $\le 3 \sigma$ agreement with the observed values from \citet{2018AJ....155..106M} (see \S\ref{sec:chi2}). In the Lyapunov time plot, simulations which did not catch into the Laplace resonance by the end of the damping phase are not shown (see \S\ref{sec:chaos}). }
    \label{fig:sum}
\end{figure*}

To efficiently sample the 3D parameter space of $K$, $\tau_a$, and $\tau_{e,1}$, we adopt the techniques of adaptive mesh refinement \citep[AMR;][]{1989JCoPh..82...64B} to focus most of our computational efforts on parts of parameter space where the system is in the Laplace resonance.
The "mesh" used here is the discretized 3D parameter space of $(K,\tau_a,\tau_{e,1})$, where at each point on the mesh we run an N-body simulation.
We apply the refinement procedure of \citet{Lohner:1987hq} \citep[see also e.g.][]{2000ApJS..131..273F} to the RMS libration amplitude of the Laplace angle.
We define the RMS libration amplitude as,
\begin{equation}
    \langle \Phi_L \rangle = \sqrt{\frac{2}{T} \int_{t_{\rm final}-T}^{t_{\rm final}} dt\, \Phi_L^2(t)} ,
\end{equation}
where the averaging time $T=10^4 P_1$ for resonant systems\footnote{For simulations that do not catch into resonance the averaging time is $10\%$ of $\tau_a$.}.
We focus our refinement on regions where $\langle \Phi_L \rangle < 100^\circ$ and where there is a boundary between an unstable\footnote{For simplicity we define "unstable" to mean an ejection of a planet from the system or when the Hill spheres of two planets overlap.} cell and a cell in resonance.
Our adaptive grid uses up to $8$ refinement levels, with each refinement level doubling the number of points in $K$, $\tau_a$, and $\tau_{e,1}$, respectively. 
Our final grid has an effective uniform resolution of $256^3$ logarithmically spaced points in the domain $\tau_a \in [10^2, 10^6] P_1$ , $K \in [1,10^4]$, and $\tau_{e,1} \in [10^4,10^8] P_1$\footnote{Preliminary simulations ruled out regions where $\tau_a < 10^2 P_1$, $K>10^4$, and $\tau_{e,1} < 10^4 P_1$.}.
We determine the AMR grid through one set of simulations, and then repeat each simulation with randomized initial phase angles for a total of $10$ simulations per $(K,\tau_a,\tau_{e,1})$.

In Figure \ref{fig:grid} we show our AMR grid and $\langle \Phi_L \rangle$ for a slice at $\tau_{e,1} = 10^5 P_1$.
The opacity of each cell shows the instability fraction, with white cells indicating that all $10$ initial conditions for that $(K,\tau_a,\tau_{e,1})$ point went unstable.
In the end, we run a total of $\sim1.3$ million simulations, roughly two orders of magnitude less than the $\sim160$ million simulations required to have a uniform resolution of $256^3$ points over our range of $K$, $\tau_a$, and $\tau_{e,1}$.
These savings allow us to explore several different variations on the setup described in \S\ref{sec:setup}, which we detail further in \S\ref{sec:ics}.

\section{Results} \label{sec:3}

Our main results are shown in Figure \ref{fig:sum}, where we plot the $(K,\tau_a)$ distributions, integrated over $\tau_{e,1}$, of several important quantities, and in Figures \ref{fig:slice_1}-\ref{fig:slice_3}, where we show the $\tau_{e,1}$ dependence of $\Phi_L$ and $e_1$ at several slices of constant $\tau_a$ and $K$.

\subsection{Resonant angles and period ratios}

The top row of Figure \ref{fig:sum} shows the libration amplitudes of the two-body, 2:1 resonance angles, $\Phi_{1,2} = 2 \lambda_2 - \lambda_1 - \varpi_2$, $\Phi_{2,3} = 2 \lambda_3 - \lambda_2 - \varpi_2$, and $\Phi_L$.
We find that the majority of parameter space does not produce systems in the Laplace resonance.
Rather, there is a bounded resonant region where the libration amplitudes of $\Phi_{1,2}$, $\Phi_{2,3}$, and $\Phi_L$ tend to be smallest in the center and quickly rise (or become unstable) at the edges.
We argue in the next section that the instabilities at $K \lesssim 10$ are the result of the eccentricity of the outer planet increasing to the point of orbit-crossing with the middle planet, while
the instabilities at $K \gtrsim 10$ are primarily due to high eccentricities of the inner planet.
The observed libration amplitude of $\Phi_L$ is of moderate to large amplitude \citep[$\simeq 20^\circ-30^\circ$;][]{2016MNRAS.455.2484N,2018AJ....155..106M}, and so we are primarily interested in the transitional regions of parameter space.

The bottom row of Figure \ref{fig:sum} shows the period ratios for the inner and outer pair of planets.
Since we start the outer planet well outside the 3:1 MMR, the outer two planets catch into MMRs with period ratios greater than 2 for slow enough damping timescales.
This is not the case for the inner pair of planets, as they are initially placed just wide of the 2:1 resonance.
Moreover, once the outer pair is locked into resonance, the effective migration rate slows by a factor dependent on the mass ratios \citep{2002ApJ...567..596L}.
This reduction of the effective migration rate makes it easier for the inner pair of planets to capture into the 2:1 resonance, rather than push through to the 3:2 MMR \citep{2014AJ....147...32G}.

\begin{figure}[h]
    \centering
    \plotone{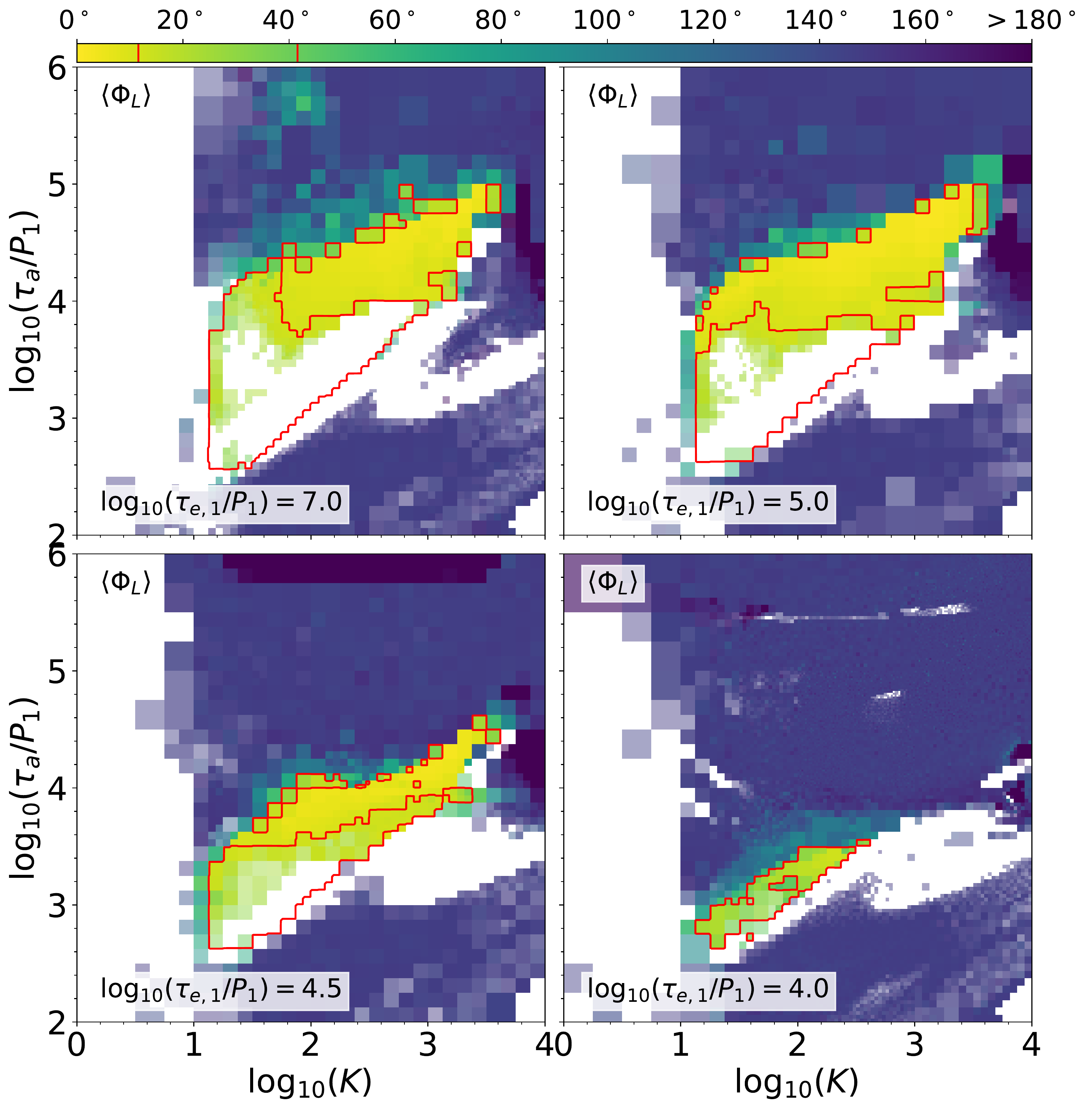}
    \caption{Similar to Figure \ref{fig:sum}, but we focus on the Laplace angle libration amplitude for four constant values of $\tau_{e,1}$: $10^{4},10^{4.5}, 10^{5}$ and $10^7 P_1$. As $\tau_{e,1}$ decreases, the resonant region shrinks and moves to lower $\tau_a$ and $K$.}
    \label{fig:slice_1}
\end{figure}

\begin{figure}[h]
    \centering
    \plotone{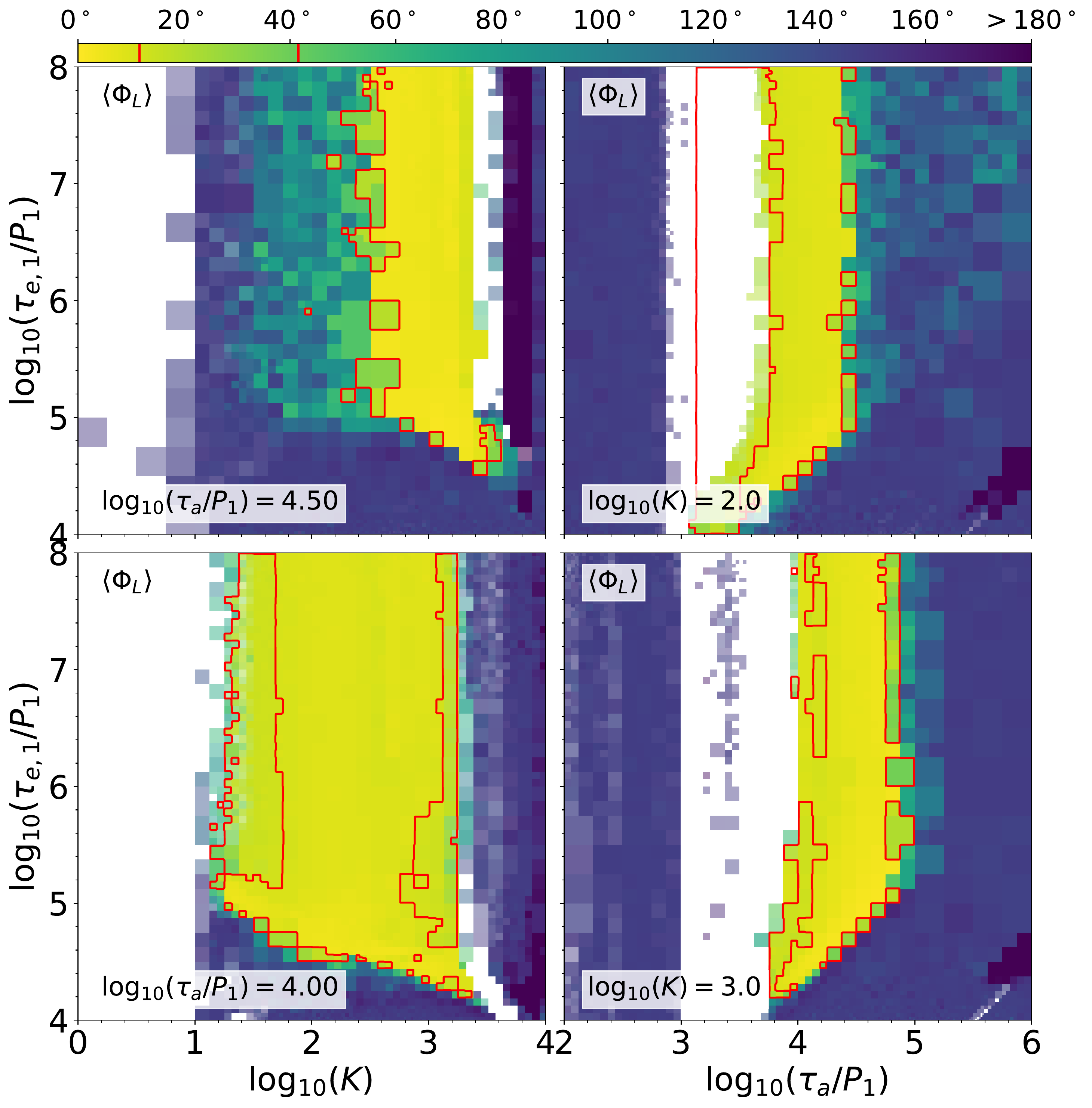}
    \caption{Similar to Figure \ref{fig:slice_1}, but for slices through parameter space for constant $\tau_a$ (left column) and constant $K$ (right column) for the Laplace angle libration amplitude. The $\tau_a$ slices are $10^{4} P_1$ and $10^{4.5} P_1$, while the $K$ slices are at $10^2$ and $10^3$.}
    \label{fig:slice_2}
\end{figure}

In Figure \ref{fig:slice_1} we show the libration amplitude of $\Phi_L$ for four different values of $\tau_{e,1}$: $10^{4} P_1$, $10^{4.5} P_1$, $10^5 P_1$, and $10^7 P_1$.
As $\tau_{e,1}$ decreases the stable Laplace resonance region shrinks and shifts towards lower $\tau_a$ and lower $K$ which were unstable at larger values of $\tau_{e,1}$.   
The magnitude of $\langle \Phi_L \rangle$ is also typically larger for lower $\tau_{e,1}$. 
Note that the unstable region for $K \lesssim 10$ is relatively robust to the value of $\tau_{e,1}$, reinforcing the understanding that this is connected to the large eccentricity of the outer planet. 
On the other hand, the region of instability at the lower end of the resonant region for $K \gtrsim 10$ shrinks as $\tau_{e,1}$ decreases, as the equilibrium eccentricity of the inner planet is lower (see also Figure \ref{fig:slice_3}).

Similar to Figure \ref{fig:slice_1}, Figure \ref{fig:slice_2} shows $\langle \Phi_L \rangle$ for cuts of constant $\tau_a$ and $K$.
In the left panels, we show the $(K,\tau_{e,1})$ dependence of $\langle \Phi_L \rangle$ for $\tau_a = 10^4 P_1$ and $10^{4.5} P_1$, while in the right panels we show the $(\tau_a,\tau_{e,1})$ dependence for $K = 10^2$ and $10^3$. 
In all cases, $\langle \Phi_L \rangle$ is roughly independent of $\tau_{e,1}$ until $\tau_{e,1} \sim \tau_a$ at which point the system quickly escapes from the Laplace resonance.

\subsection{Eccentricities}

\begin{figure}[h]
    \centering
    \plotone{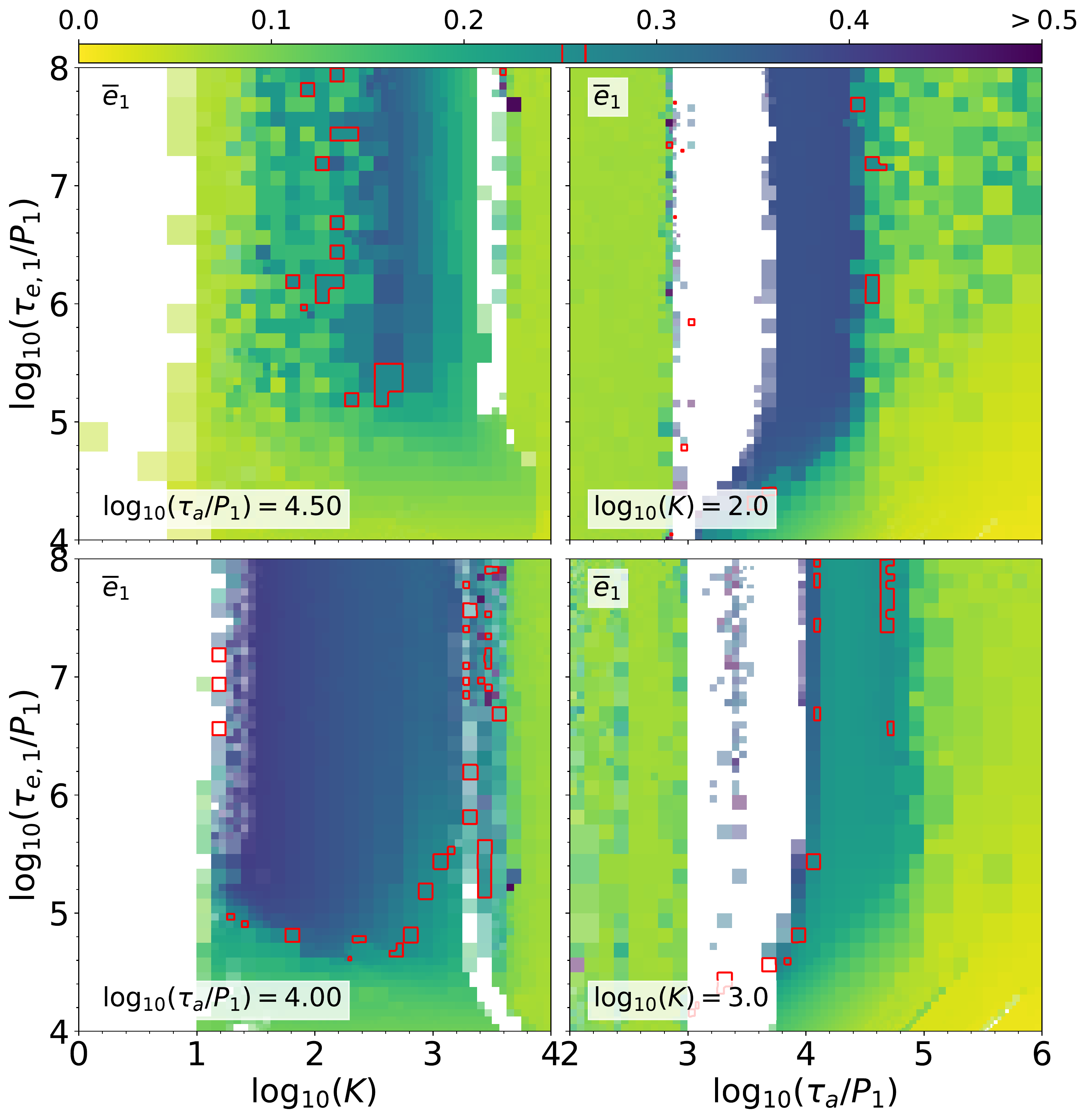}
    \caption{Same as Figure \ref{fig:slice_2}, but for the eccentricity of the inner planet.}
    \label{fig:slice_3}
\end{figure}

The middle row of Figure \ref{fig:sum} shows the final time-averaged eccentricities of each simulation.
When in the three-body resonance, the inner planet's eccentricity is typically pumped to large values ($e_1 \sim0.4$ for large $\tau_{e,1}$), while $e_2$ remains low, and $e_3$ increases as $K$ decreases.
Figure \ref{fig:slice_3} shows the $\tau_{e,1}$ dependence of $e_1$ for the same constant values of $\tau_a$ and $K$ as discussed previously in Figure \ref{fig:slice_2}. 
Similar to $\langle \Phi_L \rangle$, for fixed $\tau_a$ and $K$, the inner planet's eccentricity is roughly independent of $\tau_{e,1}$ when $\tau_{e,1}$ is large. 
When $\tau_{e,1}$ becomes comparable in magnitude to $\tau_a$ the inner planet reaches an equilibrium eccentricity lower than $e_1 \sim 0.4$. 
However, the transition to $e_1 < 0.2$ is relatively quick as $\tau_{e,1}$ decreases while the transition is more gradual for fixed $\tau_{e,1}$ and increasing either $\tau_a$ or $K$.

In the next section, we explore more quantitatively how well our simulated systems compare to the observed GJ876.

\subsection{Comparison to the real GJ876} \label{sec:chi2}

\begin{figure}[t]
    \centering
    \plotone{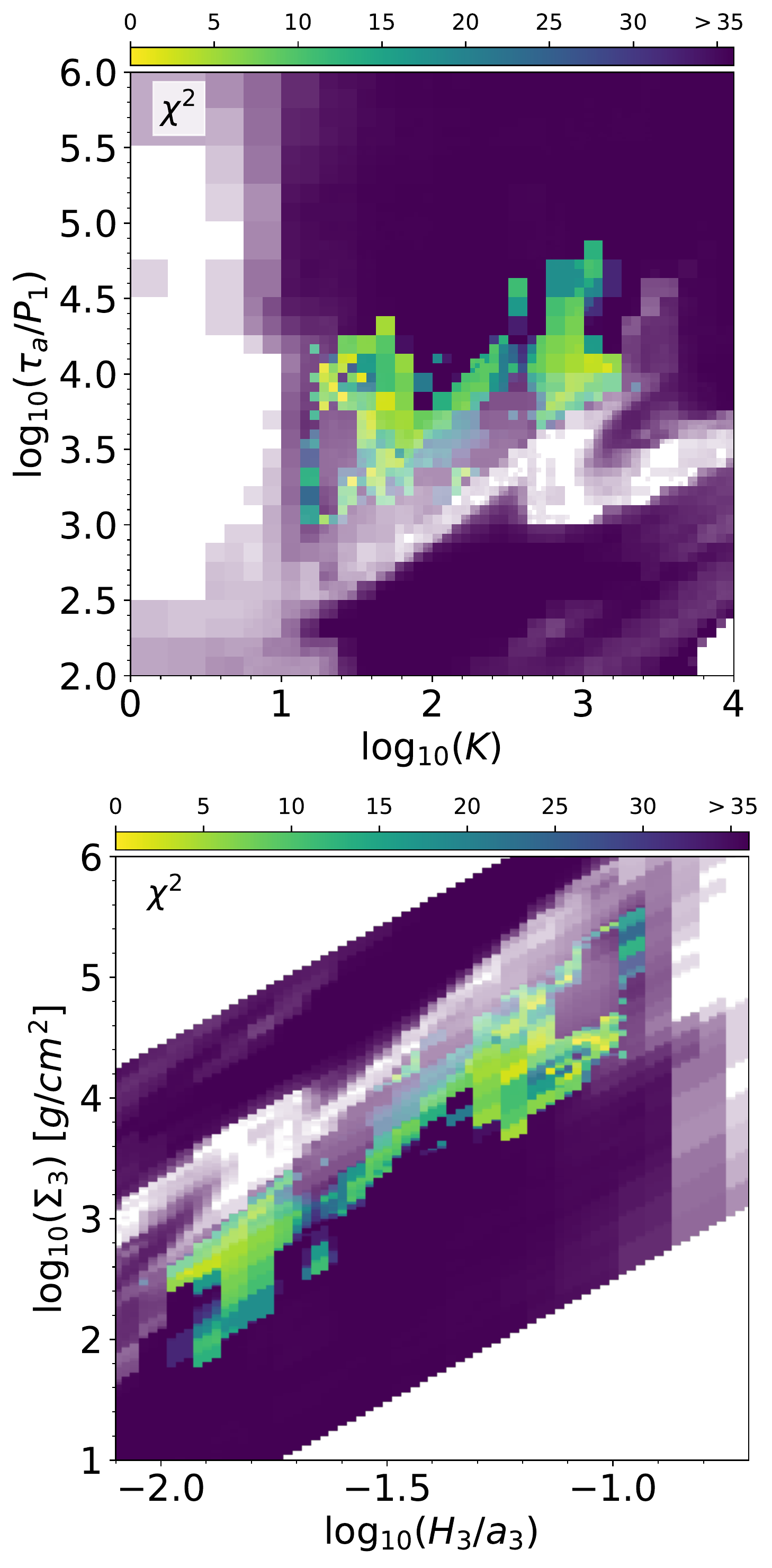}
    \caption{The $\chi^2$ statistic defined in Equation \eqref{eq:chi2}. The top panel shows $\chi^2$ as a function of $K$ and $\tau_a$. The bottom panel shows $\chi^2$ as a function of disk thickness, $H/a$, and disk surface density, $\Sigma$ at the location of the outer planet. To convert from $\tau_a$ and $K$ to $\Sigma$ and $H/a$ we use Equations \eqref{eq:tau_a} and \eqref{eq:tau_e}.}
    \label{fig:chi2}
\end{figure}

The red contours in Figures \ref{fig:sum}-\ref{fig:slice_3} show the $3 \sigma$ contours of the observed values for the three resonant angles, $\Phi_{1,2} = 10.4^\circ \pm 1.8^\circ$, $\Phi_{2,3} = 25.0^\circ \pm 4.65^\circ$, $\Phi_L = 26.6^\circ \pm 5.0^\circ$, and three eccentricities, $e_1 = 0.257 \pm 0.002$, $e_2 = 0.033 \pm 0.002$, and $e_3 = 0.03 \pm 0.023$, taken from \citet{2018AJ....155..106M}.
For simplicity, we have averaged any asymmetric errors.

The resonant angles are within the $3 \sigma$ contours for a wide range of $K$,  $\tau_a$, and $\tau_{e,1}$.
The eccentricities of the inner planets, on the other hand, are near their observed values in a much narrower range of parameter space. 
This suggests that the inner planets' eccentricities provide a strong constraint on the region of parameter space consistent with the observed system.

To quantify how well a system agrees with GJ876 we calculate a $\chi^2$ statistic for each simulation, which we define as,
\begin{equation} \label{eq:chi2}
    \chi^2 = \frac{1}{6} \sum_{i=1}^6 \left( \frac{ y_i - y_\text{obs,i}}{\sigma_\text{obs,i}} \right)^2 ,
\end{equation}
and where  $y_{\rm obs}$ and $\sigma_{\rm obs}$ are the observed values and uncertainties of $\Phi_{1,2},\Phi_{2,3}, \Phi_L, e_1, e_2$, and $e_3$.
For a given $(K,\tau_a)$ pair we show the minimum $\chi^2$ across all $\tau_{e,1}$ values in the top panel of Figure \ref{fig:chi2}.

This combined statistic attains its minimum in the region of parameter space near $K\simeq 30-3,000$ and $\tau_a\simeq 10^3 -10^4 P_1$. 
In particular, there are two regions of best fit: one around $K\simeq30$ and $\tau_a \simeq 3 \times 10^3-10^4 P_1$ and another around $K\simeq 300-1,000$ and $\tau_a \simeq 10^4 P_1$.
The region at larger $K$ and longer $\tau_a$ corresponds to systems with weak eccentricity damping on the inner planets (large $\tau_{e,1}$), while the region at lower $K$ and shorter $\tau_a$ corresponds to systems with short $\tau_{e,1}$. 

For simplicity we use equal weighting for the six parameters when computing $\chi^2$ in Equation \eqref{eq:chi2}. 
If the resonant angles are given more weight in the sum, the $\chi^2$ values generally come down since the relative uncertainties in the angles are larger, but the overall best-fit region does not change. 
This is clear from the top row of Figure \ref{fig:sum} which shows that the region enclosed by the $3 \sigma$ contours is the same for the three resonant angles and coincides with the overall low $\chi^2$ region in the top panel of Figure \ref{fig:chi2}.
Similarly, if we give more weight to the eccentricities, the minimum $\chi^2$ region as shown in Figure \ref{fig:chi2} experiences little change.

In the bottom panel of Figure \ref{fig:chi2} we use Equations \eqref{eq:tau_a} and \eqref{eq:tau_e} to derive a constraint on the surface density of gas and disk thickness near the outer Laplace planet.
Again, there are two regions of parameter space where the simulations match the observed system particularly well.
One has low surface density and thickness with $\Sigma \simeq \text{few} \times 10^2$  $\text{g/cm}^2$ and $H/a \simeq 0.01-0.02$, and the other has a higher surface density and thickness with $\Sigma \simeq \text{few} \times 10^3 - 10^5$ $\text{g/cm}^2$ and $H/a \simeq \text{few} \times 10^{-2}$. 
This latter region typically requires stronger eccentricity damping of the inner planets in order to drive the system into the observed configuration.
In other words, if we did not damp the inner planets  in addition to the outermost planet, we would infer a very thin, low surface density disk near GJ876-e at the formation time of the Laplace resonance.

As previously mentioned, these estimates depend on the overall normalizations connecting the disk properties to the semi-major axis and eccentricity damping timescales, and are best determined by hydrodynamical models \citep{1980ApJ...241..425G,1981ApJ...243.1062G,1993ApJ...419..155A,1993ApJ...419..166A,2002ApJ...567..596L,2002ApJ...565.1257T,2012ARA&A..50..211K}.
Nevertheless, it is encouraging that we obtain reasonable numbers \citep[cf.][]{2005ApJ...631.1134A} for the inner regions of GJ876's protoplanetary disk given our simplified migration model.

\subsection{Chaos} \label{sec:chaos}

\begin{figure}

    \centering
    \plotone{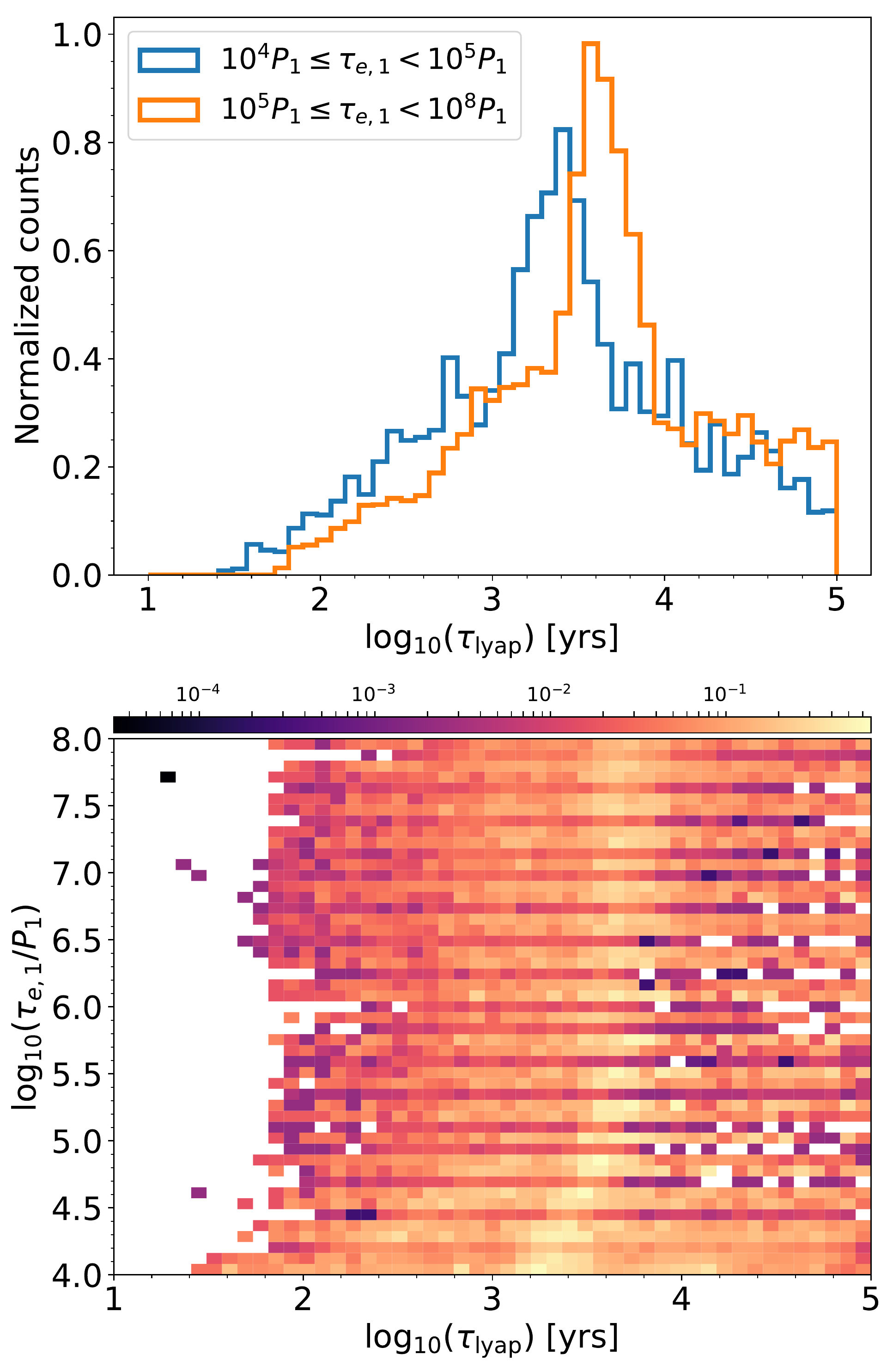}
    \caption{\emph{Top}: Normalized histogram of $\tau_{\rm lyap}$ for simulations with $10^4 \le \tau_{e,1} < 10^5 P_1$ (blue line) and $10^5 \le \tau_{e,1} < 10^8 P_1$. Because of our non-uniform grid, each simulation is given a weight of $8^{8 - \ell}$, where $\ell$ is the level of refinement for that simulation. \emph{Bottom}: A 2D histogram showing the dependence of $\tau_{\rm lyap}$ on the inner planets' eccentricity damping rate. The color of a pixel centered on a given $(\tau_{\rm lyap},\tau_{e,1})$ shows the total number of times across all $(K,\tau_a)$ pairs that the Lyapunov time was that value. The counts are normalized to the total number.}
    \label{fig:lyap_2d}
\end{figure}

Dynamical fits to GJ876 strongly suggest that the system is chaotic with a Lyapunov timescale between tens to thousands of years \citep{2015AJ....149..167B,2016MNRAS.455.2484N,Marti:2016ep,2018AJ....155..106M}.
The Lyapunov timescale, $\tau_{\rm lyap}$, characterizes the time it takes for nearly identical initial conditions to diverge, and hence is one measure of chaos \citep[e.g.][]{1999ssd..book.....M}. 
During the non-damping stage for systems deemed in the Laplace resonance, we track $\tau_{\rm lyap}$ by integrating the variational equations and monitoring the temporal evolution of the MEGNO number \citep[for a description of the algorithms used see e.g.;][]{2003PhyD..182..151C,Rein:2016hv}. 

The bottom right plot of Figure \ref{fig:sum} shows the 2D distribution of $\tau_{\rm lyap}$ for our standard set of simulations.
In the resonant region, we typically find $\tau_{\rm lyap} \gtrsim 10^3$ years, with shorter timescales on the fringes of the region.

In Figure \ref{fig:lyap_2d} we show how the Lyapunov time varies with $\tau_{e,1}$.
In the top panel, we show the distribution of $\tau_{\rm lyap}$ for all of our simulations.
Recall that we only track the chaos indicators in our simulations if they exit the damping stage with $\langle \Phi_L \rangle \le 100^\circ$. 
To get a sense of the dependence on $\tau_{e,1}$, we split the distribution into systems with strong ($\tau_{e,1} < 10^5 P_1$) and weak  ($\tau_{e,1} \ge 10^5 P_1$)  eccentricity damping on the inner planets. 
The distributions are mostly similar with peaks around a few thousand years, but systems with shorter $\tau_{e,1}$ have a higher chance of having $\tau_{\rm lyap} < 10^3$ years. 
We explore this further in the bottom panel of Figure \ref{fig:lyap_2d}, where we show the 2D distribution of $\tau_{\rm lyap}$ and $\tau_{e,1}$.
For a given value of $\tau_{e,1}$ we compute the histogram  of $\tau_{\rm lyap}$ across all values of $K$ and $\tau_a$.
The color of each pixel in Figure \ref{fig:lyap_2d} shows the normalized, total number of times that particular value of $\tau_{\rm lyap}$ occurred. 
Again, the peak of the distribution for all $\tau_{e,1}$ occurs around Lyapunov times of a few thousand years.
At the strongest damping rates, the distribution flattens out with more occurrences of $\tau_{\rm lyap} < 10^3$ years and $\tau_{\rm lyap} > 10^4$ years. 

\begin{figure}
    \centering
    \plotone{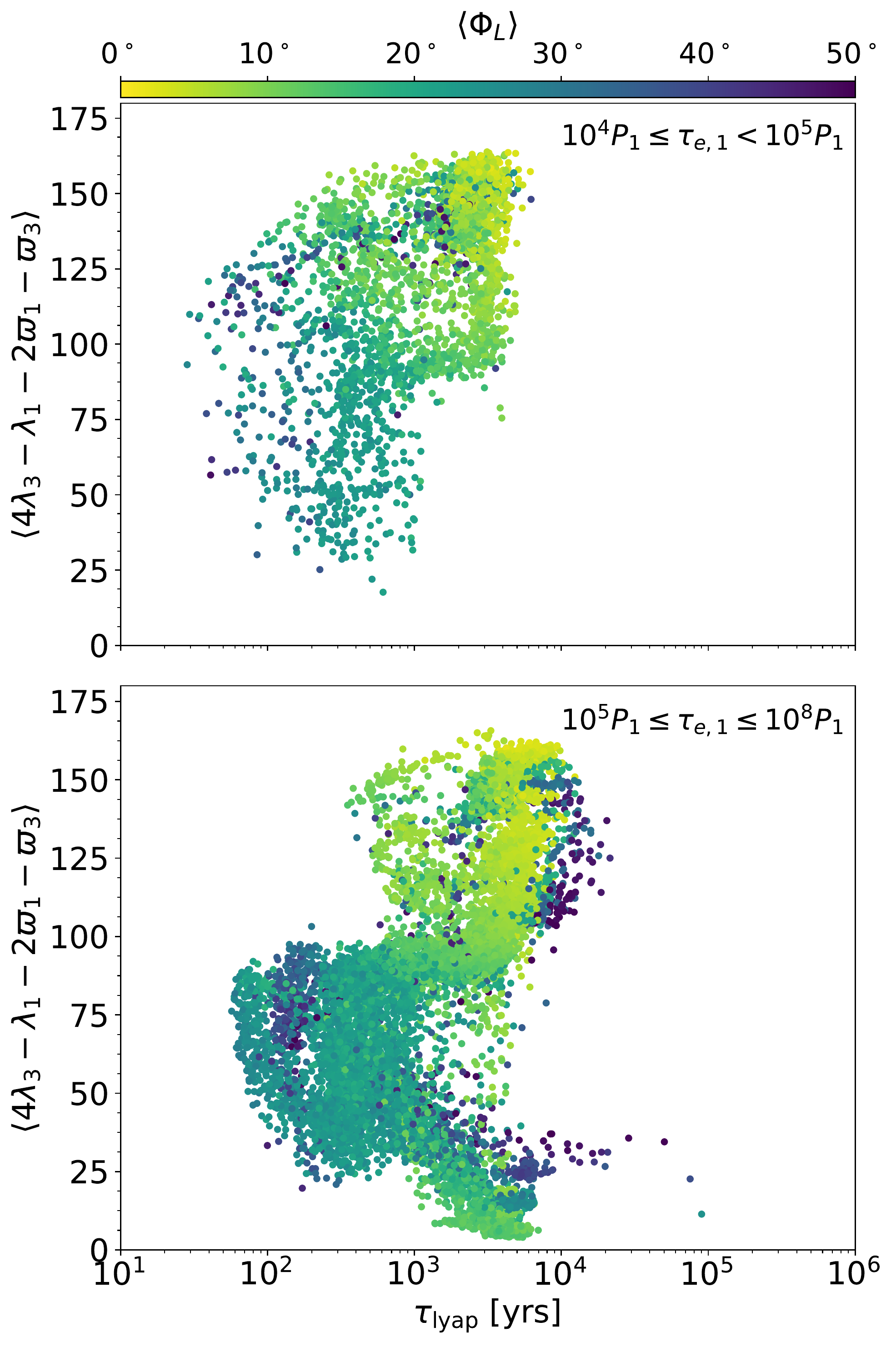}
    \caption{Correlations of the libration amplitudes for the 4:1 resonant angle, $4 \lambda_3 - \lambda_1 - 2 \varpi_1 - \varpi_3$, with $\tau_{\rm lyap}$, for two ranges of $\tau_{e,1}$.
    The top panel is for stronger eccentricity damping of the inner planet ($\tau_{e,1} < 10^5 P_1$), while the bottom panel is for weaker eccentricity damping  ($\tau_{e,1} \ge 10^5 P_1$).
    Only simulations with $\langle \Phi_L \rangle \le 50^\circ$ are shown and the color of each point corresponds to the value of $
    \langle \Phi_L\rangle$. Unlike for the Laplace angle, the libration amplitude here corresponds to the standard deviation (as opposed to the RMS).}
    \label{fig:aqr}
\end{figure}

\subsubsection{The double apsidal corotation resonance}

A longer $\tau_{\rm lyap}$ is consistent with \citet{2015AJ....149..167B}, who found that in addition to the decades long $\tau_{\rm lyap}$, there was also a region of phase space where $\tau_{\rm lyap} \gtrsim 10^3$ years. 
Moreover, \citet{2018AJ....155..106M} associate this with a low energy, double apsidal corotation resonance where the angles $\varpi_3-\varpi_2$, $\varpi_2-\varpi_1$, $2 \lambda_3 - \lambda_2 - \varpi_3$, and $4 \lambda_3 - \lambda_1 - 2 \varpi_1 - \varpi_3$ all librate.
Figure \ref{fig:aqr} shows the correlation of one of these angles, $4 \lambda_3 - \lambda_1 - 2 \varpi_1 - \varpi_3$, with $\tau_{\rm lyap}$ and $\tau_{e,1}$ for simulations with $\langle \Phi_L  \rangle < 50^\circ$\footnote{The other angles associated with the double apsidal corotation resonance show the same correlation with $\tau_{\rm lyap}$ and $\tau_{e,1}$.}.
This angle is associated with the 4:1 MMR between the innermost and outermost planets of the resonant chain. 
The top panel shows the distribution for $\tau_{e,1} < 10^5 P_1$, while the bottom panel shows the distribution for $\tau_{e,1} \ge 10^5 P_1$.
Systems with $\tau_{\rm lyap} \gtrsim 10^3$ years can either have these angles librate with small amplitude or circulate and have $\langle \Phi_L \rangle \lesssim 10^\circ$, while systems with $\tau_{\rm lyap} \lesssim 10^3$ years have predominantly larger libration amplitudes and $\langle \Phi_L \rangle \gtrsim 10^\circ$. 
This large libration amplitude suggests that the angles are switching between libration and circulation as pointed out in \citet{2018AJ....155..106M}. 

Moreover, for the shortest $\tau_{e,1}$ values, we do not find any simulations where these angles have low librations amplitudes.
This suggests that if the true system lies in the region of parameter space where the angles associated with the double apsidal corotation resonance librate with small amplitude, then there was little or no eccentricity damping of the inner planets at the time of the Laplace resonance formation.

\section{Discussion} \label{sec:4}

\subsection{Alternative initial conditions} \label{sec:ics}

\begin{figure}
    \centering
    \plotone{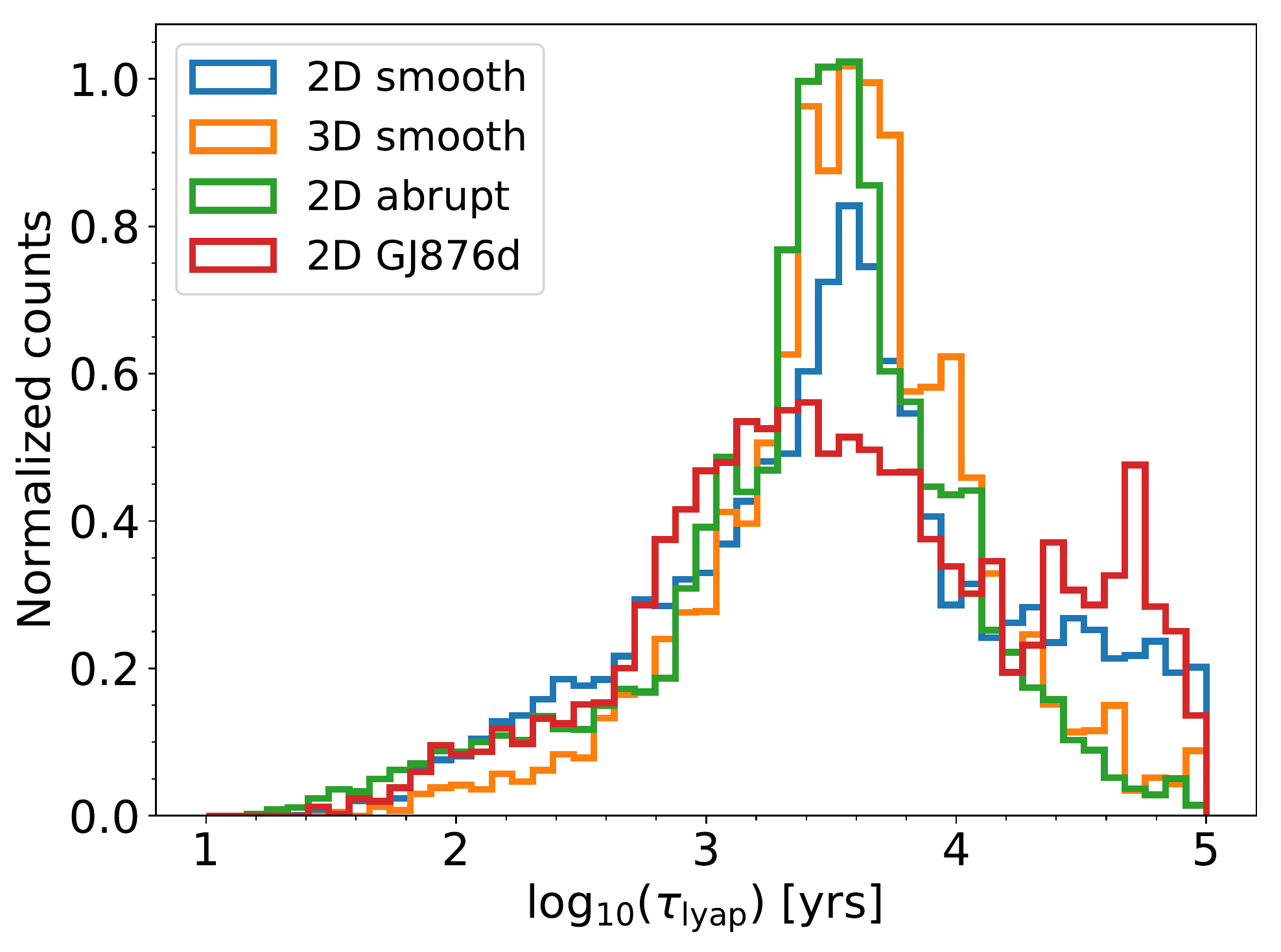}
    \plotone{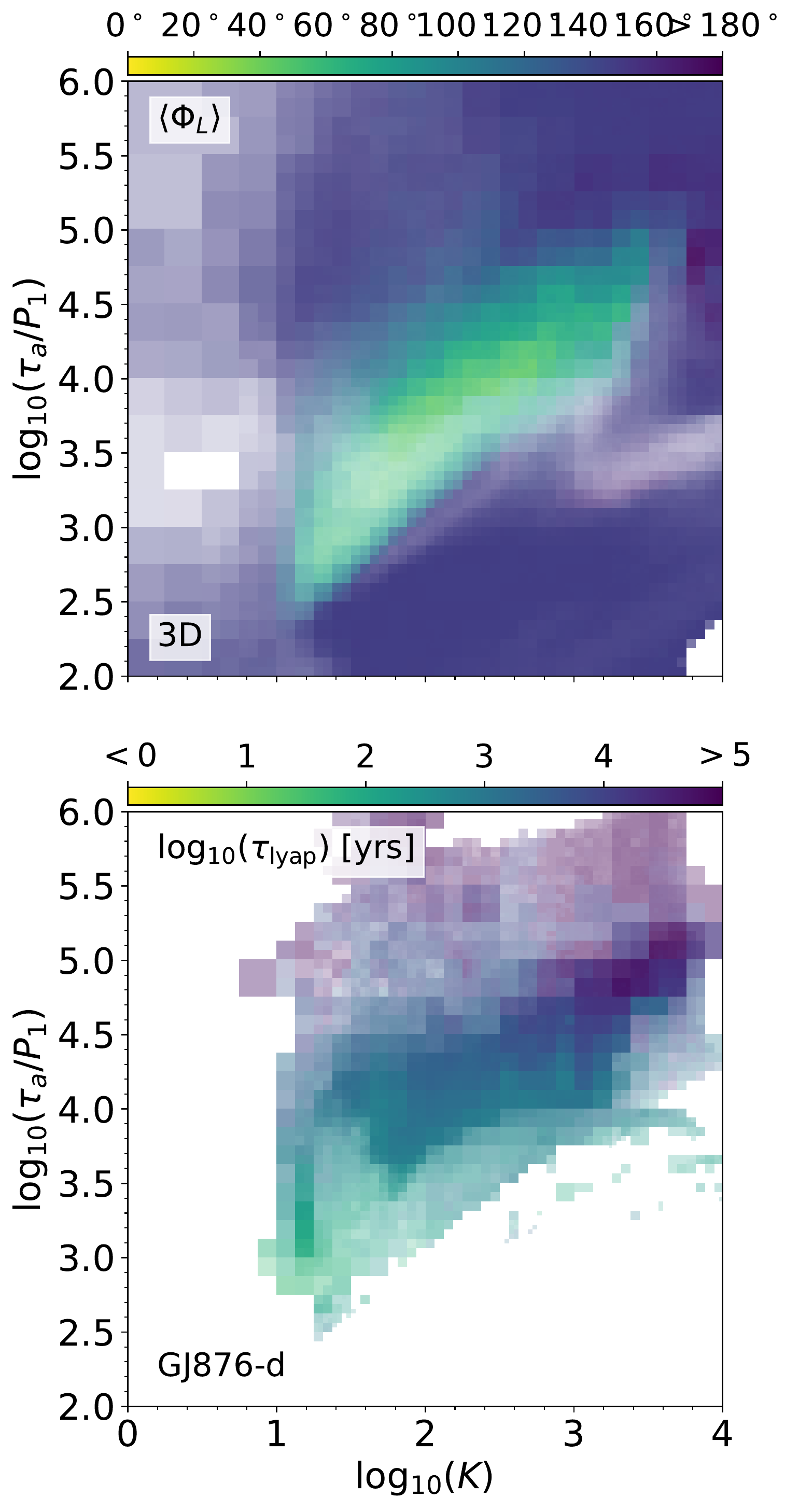}
    \caption{Results from variations on our initial setup. \emph{Top}: The distribution of $\tau_{\rm lyap}$ for the different setups described in \S\ref{sec:ics}. 
    \emph{Middle}: The $\tau_{e,1}$ integrated $\langle \Phi_L \rangle$ for the 3D setup.
     \emph{Bottom}: The $\tau_{e,1}$ integrated $\tau_{\rm lyap}$ for the setup which includes GJ876-d.}
    \label{fig:lyap_3}
\end{figure}

Given that the majority of our simulations sit in the longer Lyapunov timescale part of phase space, we wish to now examine possible alternatives to our standard setup described in \S\ref{sec:setup} which could potentially increase the number of initial conditions leading to $\tau_{\rm lyap} \lesssim 10^3$ years.
In Figure \ref{fig:lyap_3} we summarize the results of relaxing three of our initial assumptions: the timescale of the damping removal, coplanarity, and neglecting the fourth planet of the system, GJ876-d. 

\paragraph{Abrupt damping removal}
In our standard set of simulations, we slowly removed the damping forces on the outer planet over one damping timescale.
We ran an additional set of simulations where we instead abruptly turned off all damping forces.
The distribution of $\tau_{\rm lyap}$ as shown in the top panel of Figure \ref{fig:lyap_3} is relatively insensitive to how smoothly the damping forces are removed.
The only significant change is slightly more systems to the left of the main peak at $\tau_{\rm lyap} \simeq 3000$ years.

\paragraph{Mutual inclinations}

Another major simplification of our study is in assuming that the Laplace planets are coplanar with each other.
This is a relatively robust assumption, however, since the observed Laplace planets are nearly coplanar \citep{2016MNRAS.455.2484N}.
Nevertheless, we ran an additional set of simulations where we gave each planet a random initial inclination of up to $20^\circ$.
The distribution of $\tau_{\rm lyap}$ is relatively unaffected by allowing the planets to be mutually inclined.
The middle panel of Figure \ref{fig:lyap_3} shows that allowing the planets to have non-zero inclinations narrows the Laplace resonance parameter space and increases the instability fractions.

\paragraph{GJ876-d}
The two-day inner planet, GJ876-d ($M_d \simeq 7 M_\oplus,  P_d = 1.94 d, e_d \simeq 0.11$\footnote{We use the larger eccentricity from \citet{2016MNRAS.455.2484N} as opposed to the smaller eccentricity from \citet{2018AJ....155..106M}.}), has so far been neglected in the resonance capture and subsequent evolution of our models \citep{2016MNRAS.455.2484N}.
While this is likely a good assumption for the process of resonance capture, it is less so when studying the overall system parameters (such as instability and chaos) during the long term evolution of the entire system.
To this end, we ran our standard set of simulations again, but after removing the damping forces we additionally included GJ876-d at its observed period and eccentricity.
When GJ876-d is included, we also include the precessional effects of GR via REBOUNDX \citep{1986IAUS..114..105N}.
Including GJ876-d does not noticeably affect the distribution for $\tau_{\rm lyap} < 10^3$ years. 
Surprisingly, however, it does increase the frequency of systems with $\tau_{\rm lyap} > 10^4$ years.

Additionally, other effects not present in our simplified smooth migration models, such as the stochastic forcing used in \citet{2015AJ....149..167B}, or the presence of an eccentric disk during the damping phase \citep[e.g.][]{2005A&A...437..727K}, may shorten or significantly affect the chaotic timescale of the system.

\section{Conclusions and future work}\label{sec:5}

Since the majority of parameter space where GJ876 is in resonance has $\langle \Phi_L \rangle \lesssim 10^\circ$ and $\tau_{\rm lyap} \gtrsim 10^{3}$ years, it is somewhat surprising that the observed system has such large libration amplitude and short Lyapunov time.
Further observations of the system will yield better constraints on these two characteristics as well as on the eccentricities of the planets \citep{2018AJ....155..106M}.
In particular, the eccentricity of GJ876-c provides the strongest constraint on the appropriate damping timescales driving the system into resonance.

In addition to further observations, more detailed, hydrodynamical simulations of the three Laplace planets in GJ876 are required to determine whether or not the formation scenario we have presented here is truly viable.
In particular, such simulations can help explain how an eccentric disk \citep{2005A&A...437..727K}, the presence of an inner disk \citep{2008A&A...483..325C}, how the disk disperses \citep{2005A&A...437..727K}, the level of disk viscosity \citep[e.g.][]{2017ApJ...839..100F}, and other effects determine the migration rates, final eccentricities, resonance angles, and Lyapunov timescale of GJ876.
Hydrodynamical simulations will also improve upon the connection between damping timescales and disk properties (e.g. Equations \eqref{eq:tau_a} and \eqref{eq:tau_e}).

On this topic, \citet{2018arXiv180804223C} have recently simulated the construction of the Laplace resonance in GJ876 for several different disk thicknesses, masses, and viscosities using hydrodynamical simulations of two-dimensional locally isothermal disks.
Compared to our smooth migration model, the migration rate of GJ876-e in their simulations can vary wildly with time as the presence of an eccentric disk outside of GJ876-c,b induces large variations in the torque felt by the outer planet.
In many of their cases this prevents the system from forming the 4:2:1 MMR as GJ876-e is caught at larger period ratios with GJ876-b.  
In order to form the Laplace resonance, they offer two main avenues.
The first is for GJ876-e to open a partial gap in the disk, allowing it to gradually remove the disk eccentricity induced by GJ876-b,c and migrate into the 2:1 resonance with GJ876-b. 
This scenario requires the disk thickness and viscosity to be low enough for GJ876-e to open a (partial) gap.  
The second is for the resonance capture to take place later in the systems lifetime when the disk surface density is lower. 
Further simulations covering a wider range of thicknesses and disk masses would facilitate a more robust comparison to our results (e.g.\,to Figure \ref{fig:chi2}).

A natural extension of our AMR method is to the study of other resonant chain systems. 
While GJ876 is unique in its diversity of masses, systems discovered by \emph{Kepler} typically have roughly equal mass planets orbiting their host star in a compact configuration. 
Our method can be readily applied to systems containing resonant chains of Earth mass, or Super-Earth mass planets, and can put meaningful constraints on the local disk properties present when the planets captured into resonance \citep[see also e.g.][]{2015A&A...579A.128D}.
Examples of such systems include Kepler-60, Kepler-223, Trappist 1, and others \citep{2016Natur.533..509M,2016MNRAS.455L.104G,2017NatAs...1E.129L}.
In particular, chains of non-gap opening planets may undergo Type I migration, and if they have measured masses, the parameter space to explore consists only of the density and temperature profile of the disk.
This low-dimensional parameter space allows for our efficient AMR scheme to fully explore the relevant parameter space.

\acknowledgments
The mesh refinement code used is available at \url{http://github.com/adamdempsey90/NDTAMR}.
We would like to thank the referee for a thorough and insightful report which greatly improved the manuscript.
We would also like to thank Sam Hadden, Yoram Lithwick, Diego Mu{\~n}oz, Matt Payne, Hanno Rein, and Dan Tamayo for useful discussions, as well as Sarah Millholland and Greg Laughlin for sharing an early version of their paper with us.
This research was supported in part through the computational resources and staff contributions provided for the Quest high performance computing facility at Northwestern University which is jointly supported by the Office of the Provost, the Office for Research, and Northwestern University Information Technology. 
B.E.N. acknowledges support from the Data Science Initiative at Northwestern University.

\end{document}